%% file: regular_csp_hardness.tex
\documentclass[
10pt, 
a4paper, 
]{article}

\input{structure.tex} 

\hypersetup{
	colorlinks,
	linkcolor={red!50!black},
	citecolor={blue!50!black},
	urlcolor={blue!80!black}
}
\usepackage{color,soul}
\title{On Regularity of Max-CSPs and Min-CSPs} 
\author{%
  {Aleksa Stankovi\'c \thanks{Research supported by the Approximability and Proof Complexity project funded by the Knut and Alice Wallenberg Foundation.}} \\[1ex] 
\normalsize KTH Royal Institute of Technology \\ 
\normalsize \href{mailto:aleksas@aleksas.se}{aleksas@kth.se} 
}
\date{\today} 

\DeclareMathOperator{\Pal}{P}
\DeclareMathOperator{\Prob}{P}
\DeclareMathOperator{\NP}{NP}
\DeclareMathOperator{\ar}{ar}
\DeclareMathOperator{\Opt}{Opt}
\DeclareMathOperator{\Val}{Val}
\DeclareMathOperator{\f1}{\mathbf{1}}
\DeclareMathOperator{\E}{\mathbf{E}}
\newcommand*\diff{\mathop{}\!\mathrm{d}}
\definecolor{NavyBlue}{rgb}{0, 0, 0}
\begin{document}

\maketitle

\begin{abstract}
\noindent 
We study approximability of regular constraint satisfaction problems, i.e., CSPs where each variable in an instance has the same number of occurrences. In particular, we show that for any CSP $\Lambda$, existence of an $\alpha$ approximation algorithm for unweighted regular Max-CSP $\Lambda$ implies existence of an $\alpha-o(1)$ approximation algorithm for weighted Max-CSP $\Lambda$ in which regularity of the instances is not imposed. We also give an analogous result for Min-CSPs, and therefore show that up to arbitrarily small error it is sufficient to conduct the study of approximability of CSPs only on regular unweighted instances. 
\end{abstract}


\section{Introduction}
		This work studies approximability of regular constraint satisfaction problems (CSPs), where we interpret regularity to mean that each variable appears the same number of times in constraints of instances. Since regular CSPs are a subclass of CSPs, approximating their optimal values is not harder than approximating values of general CSPs. In this work we show that approximating values of regular CSPs is also \emph{essentially} not easier, i.e. we show that an $\alpha$ approximation algorithm for regular instances of a particular CSP induces $\alpha-o(1)$ approximation algorithm applicable to possibly non-regular instances. Therefore, we show that imposing regularity has \emph{almost} no effect on the approximability of CSPs, and in particular if one is not interested in $o(1)$ additive factors in approximation ratios, the study of approximability may be conducted solely on regular instances. 
\par In order to make the result more general, we revisit previously studied question of \emph{weights vs. no weights} for CSPs \cite{DBLP:conf/istcs/CrescenziST96,DBLP:journals/siamcomp/KhannaSTW00} in a context of approximation. In particular, we show that it is sufficient to have an $\alpha$ approximation algorithm for regular \emph{unweighted} instances in order to construct an $\alpha - o(1)$ approximation algorithm applicable to possibly \emph{weighted} instances of CSPs without regularity restriction. In order to do so, we use a result from \cite{DBLP:journals/siamcomp/KhannaSTW00} which shows that weighted versions of CSPs have \emph{essentially} the same (up to $o(1)$ additive error) approximation ratios as their unweighted counterparts. We reprove this result here for the sake of completeness.
\par We organize the paper as follows. In Section \ref{section_CSP}, we give an informal definition of constraint satisfaction problems, and introduce decision and optimization versions of these problems. In Section \ref{section_approx_CSP}, we discuss approximation of CSPs and highlight some breakthrough results. Motivated by this discussion, we introduce regular CSPs in Section \ref{section_1_results}, and state the new results proved in this work. Then, in Section \ref{section_1_comparison} we compare the results of this paper with prior work.
		In Sections \ref{prelim_section} and \ref{red_section} we formalize the discussion given in Section 1. In particular, in Section \ref{prelim_section} we fix the notation, and discuss the difference between weighted and unweighted CSPs. In Section \ref{red_section} we describe our reductions and prove the results. Finally, in Section \ref{conclusion_section} we discuss possible applications of ideas and theorems introduced in this paper, and mention some open questions.
		\subsection{Constraint Satisfaction Problems}		\label{section_CSP}

    Constraint satisfaction problems (CSPs) represent one of the most fundamental classes of problems studied in complexity theory. Each CSP is described by a collection of predicates, which are used in instances of these problems as constraints on tuples of variables. Probably the best known CSP is 3-Sat, in which the constraints are given as disjunctive clauses on at most three literals, where a literal is either a variable or its negation. A basic problem is to determine whether we can satisfy all the constraints of a given CSP instance. 
		
		This problem is very well understood, due to Schaefer's dichotomy theorem for CSPs on Boolean domains \cite{Schaefer:1978:CSP:800133.804350} and more recent proofs of a dichotomy theorem on general domains by Bulatov \cite{DBLP:conf/focs/Bulatov17} and Zhuk \cite{DBLP:conf/focs/Zhuk17}.
    \par 
				In this work we focus on optimization variants of CSPs, in which we are interested in finding an assignment which maximizes/minimizes the number of constraints satisfied. Depending on the optimization variant, we refer to these problems as either Max-CSPs or Min-CSPs. A typical problem in this setting is Max-Cut, which has Boolean constraints of the form $x_i \neq x_j$. Many optimization CSPs are intractable, and in this case we typically resort to approximation algorithms in order to estimate their optimal values. The strength of an approximation algorithm is expressed through its approximation ratio $\alpha$, which measures the quality of a solution produced by the algorithm by comparing\footnote{By convention we assume in this work that approximation algorithms for Max-CSPs always have $\alpha<1$, while for Min-CSPs $\alpha>1$.} it to the optimal one. In a study of the approximation algorithms, we are typically interested in finding algorithms with the value of $\alpha$ as close to $1$ as possible. We are also interested in studying which values of $\alpha$ are not feasible, in which case we talk about inapproximability. 
    \par
		\subsection{Some Important Results on Approximability of CSPs} \label{section_approx_CSP}
    On the algorithmic side, semidefinite programming (SDP) has been very fruitful tool for approximating optimal values of CSPs. The first approximation algorithm based on SDP dates back to the work of Goemans and Williamson, who devised a $\approx 0.878$ approximation algorithm for the Max-Cut problem \cite{DBLP:conf/stoc/GoemansW94}. Ideas from this work have been very influential for subsequent research of approximation algorithms, and we highlight the $7/8$ approximation algorithm for Max-3-Sat \cite{DBLP:conf/focs/KarloffZ97} and the $0.940$ algorithm for Max-$2$-Sat \cite{DBLP:conf/stoc/Austrin07,DBLP:conf/ipco/LewinLZ02}.
    \par 
     
    On the hardness of approximation side the celebrated PCP-theorem \cite{DBLP:conf/focs/AroraLMSS92, DBLP:conf/focs/AroraS92}, combined with the usual assumption that $\Pal \neq \NP$, provided a strong starting point used in many results showing impossibility of approximation. The highlight result using this starting point along with parallel repetition of Raz \cite{DBLP:journals/siamcomp/Raz98} and long codes \cite{DBLP:journals/siamcomp/BellareGS98} comes from H\r{a}stad, who gave optimal inapproximability results for Max-E$k$-Sat and Max-E$k$-Lin\footnote{In Max-E$k$-Sat, constraints are clauses of width $k$, while in Max-E$k$-Lin, constraints are linear equations over $\mathbb{Z}_2$. We use abbreviation E$k$ to denote that each constraint is of width exactly $k$. Therefore, Max-E$k$-Sat allows only clauses of width $3$, while Max-$3$-Sat allows width $1$ and $2$ as well.}  problems \cite{DBLP:journals/jacm/Hastad01}. Recently, Siu On Chan gave optimal (up to a constant factor) inapproximability results for Max-CSPs where arity $k$ of the predicates is larger than the size of the domain \cite{DBLP:journals/jacm/Chan16}. 
    \par 
		Even though the PCP theorem was used with great success over the years, researchers still faced seemingly insurmountable difficulties in pursuit of sharp inapproximability results for many fundamental problems such as Max-2-Sat, approximate graph coloring, and minimum vertex cover. More precisely, the starting point of almost all reductions was the Label Cover problem \cite{DBLP:journals/jcss/AroraBSS97}, which was constructed by combining the PCP theorem with the parallel repetition of Raz \cite{DBLP:journals/siamcomp/Raz98}. In order to overcome these difficulties, Khot introduced a modification of Label Cover called \emph{Unique Label Cover} \cite{DBLP:conf/stoc/Khot02a} and conjectured it to be NP-Hard. This conjecture is known as the Unique Games Conjecture (UGC), and it quickly became the central problem in the hardness of approximation area, especially since its validity implies optimality of many already known approximation algorithms. Of special importance among UGC based results is the one from Raghavendra \cite{DBLP:conf/stoc/Raghavendra08}, which shows that a certain version of semidefinite programming relaxation is optimal for all constraint satisfaction problems. Therefore, in case UGC is shown to be true, this work would end the quest for optimal approximation algorithms for CSPs.
    \par
    However, with the validity of the UGC still in question, there is an incentive to derive strong inapproximability results relying on other (weaker) assumptions, most preferably on $\Pal \neq \NP$. Furthermore, while Raghavendra's result shows how to optimally approximate CSPs, it does not give us a suitable way to compute numerical values of optimal approximation ratios; this question remains open for almost all CSPs, even very simple ones.  
		\subsection{New Results for Regular CSPs} \label{section_1_results}
    In order to facilitate further study, it can be valuable to ask whether some additional properties of instances can be assumed when studying approximability of CSPs. In this work we address this topic by studying regular instances of Max-CSPs and Min-CSPs, i.e.~instances in which each variable occurs the same number of times in the constraints. In particular, we prove the following results for poly-time approximation algorithms.
    \begin{theorem} \label{main_theorem_introduction_max}
			If there is an $\alpha$ approximation algorithm for unweighted regular instances of Max-CSP $\Lambda$ then for every $\delta > 0 $ there is an $\alpha-\delta$ approximation algorithm for the weighted Max-CSP $\Lambda$.
    \end{theorem}
    \begin{theorem} \label{main_theorem_introduction_min}
			If there is an $\alpha$ approximation algorithm for unweighted regular instances of Min-CSP $\Lambda$ then for every $\delta>0$ there is an $\alpha+\delta$ approximation algorithm for the weighted Min-CSP $\Lambda$.
    \end{theorem}
		The proofs of Theorems \ref{main_theorem_introduction_max} and \ref{main_theorem_introduction_min} are based on a deterministic reduction introduced in Theorem \ref{max_min_csp_det_theorem}. We also give a randomized reduction for Max-CSPs in order to prove the following theorem.
    \begin{theorem} \label{main_theorem_introduction_max_random}
			Given a Max-CSP $\Lambda$ and $\delta>0$, it is sufficient to have an $\alpha$ approximation algorithm for unweighted regular instances of degree up to $O(\log(1/\delta)/\delta^2)$ to have an $\alpha-\delta$ randomized approximation algorithm for the weighted Max-CSP $\Lambda$ with success probability of at least $1-O(2^{-n})$, where $n$ is the number of variables appearing in an instance. 
    \end{theorem}
		The details of the randomized reduction can be found in Theorem \ref{max_csp_rand_theorem}. Randomized reduction also works for Min-CSPs, although with the degree requirement of $\Omega(n^2\log(n))$, which makes this reduction less efficient than even the deterministic one. For this reason we do not discuss randomized reduction for Min-CSPs. 
		\par
			In the theorems above instead of a constant $\delta$ we can choose $\delta = \Omega(1/poly(n))$, where $n$ is the number of variables, to obtain $\alpha-o(1)$ approximation in poly-time for Max-CSPs (or $\alpha+o(1)$ approximation for Min-CSPs). \par 
    \par 
		\subsection{Prior Work} \label{section_1_comparison}	
		Both the randomized and the deterministic reductions introduced in this paper are based on a construction introduced by Trevisan \cite{DBLP:conf/stoc/Trevisan01}, which was used to show hardness of approximating values of bounded degree instances of the Max-$3$-Sat problem.  The reduction of Trevisan outputs instances in which each variable has degree $D$ in expectation, and therefore by argument that relies on Chernoff's bound it is shown that the degrees of variables is with high probability smaller than $D e^2$. Our deterministic reduction comes from derandomization of the aforementioned result, while in the randomized reduction we reuse mentioned result of Trevisan \cite{DBLP:conf/stoc/Trevisan01} and in our argument show that degrees are with high probability in range $[D-o(1),D+o(1)]$.
		\par 
		In order to make our reductions applicable to the weighted setting, in this work we also show that approximability of weighted Max-CSPs (or weighted Min-CSPs) is \emph{essentially the same}\footnote{If we allow $o(1)$ additive loss in the approximation ratio} as the approximability of their unweighted versions. Let us remark that the same result was already proved in \cite[Lemma 3.11]{DBLP:journals/siamcomp/KhannaSTW00} by relying on some results that appeared in \cite{DBLP:conf/istcs/CrescenziST96}. We reprove this fact here for the sake of completeness.


    \par 


  \section{Preliminaries}\label{prelim_section}
    We consider constraint satisfaction problems given by the following definition.
    \begin{definition}
      A \textbf{constraint satisfaction problem} (CSP) over a language $ \Sigma = [q], q \in \mathbb{N}$, is a finite collection of predicates $\Lambda \subseteq \{ P:[q] ^k \to \{0,1\} \mid k \in \mathbb{N}\}$. 
    \end{definition}
    For a predicate $P:[q] ^k \to \{0,1\}$ we use $\ar(P)=k$ to denote its arity. We are interested in solving instances of CSPs, which are defined as follows.
    \begin{definition} \label{csp_inst_def}
      An \textbf{instance} $\mathcal{F}$ of a CSP $\Lambda$ consists of a set $X = \{x_1,\hdots,x_n\}$ of $n$ variables taking values in $\Sigma$ and a set $\mathcal{C} = \{C_1,\hdots,C_{m}\}$ of $m$ constraints, where each constraint $C_r$ is a pair $(P_r,S_r)$, with $P_r \in \Lambda$ being a predicate with arity $k_r := \ar(P_r)$, and $S_r$ being an ordered tuple containing $k_r$ \emph{distinct} variables which we call a \textbf{scope}.
    \end{definition}
    Sometimes when working with Boolean CSPs, the definition of an instance allows applying constraints to literals instead of variables. However, Definition \ref{csp_inst_def} is more general, since we can always extend the family of predicates $\{P_r\}_r$ belonging to a CSP $\Lambda$ to create CSP $\overline{\Lambda}$, such that each instance of $\overline{\Lambda}$ in the sense of Definition \ref{csp_inst_def} can be represented as an instance of $\Lambda$ in which we allow constraints over variables and their negations, and vice-versa. In particular, we can create $\overline{\Lambda}$ by taking every $P_r$ of $\Lambda$, considering all $I \subseteq \{0,1\}^{\ar(P_r)}$, and adding to $\overline{\Lambda}$ predicates $P_r^I$ defined as \begin{equation*}
    P_r^I(x_1,\hdots,x_n) = P_r(x_1+I_1,x_2+I_2,\hdots,x_{\ar(P_r)}+I_{\ar(P_r)} ),
  \end{equation*}
  where $I_i$ is the $i$-th element of the tuple $I$, and addition takes place over $\mathbb{Z}_2$.


    The degree $d_i$ of a variable $x_i$ is defined as the number of times $x_i$ is mentioned in the constraints, or formally
    \begin{equation}
      d_i = \sum\limits_{r=1}^{m} \f1_{S_r}(x_i),  
    \end{equation} where $\f1_{S_r}$ is an indicator function. Instances in which all variables have the same degree are called \textbf{regular}. 
    \par 
    Max/Min-CSP problems frequently appear in a setting in which constraints of an instance are assigned with non-negative weights, which are typically used to encapsulate the significance of each constraint. Let us now give the definition of these problems.
    \begin{definition}
      A \textbf{weighted instance} $\mathcal{F}$ of a CSP $\Lambda$ is an instance of $\Lambda$, where each constraint $C_r$ has a weight $w_r \geq 0$, and $\sum_r w_r= 1$ .
    \end{definition}
    Obviously, unweighted instances can be seen as weighted where each constraint $C_r$ has a weight $w_r= 1/m$. 
    \par
    Let us denote by a function $\chi \colon X  \to \Sigma$ an assignment to variables $X$ of a CSP instance $\mathcal{F}$. We interpret $\chi(S_i)$ as a coordinate-wise action of $\chi$ on $S_i$. Given $\chi$, we define the value $\Val_{\chi}(\mathcal{F})$ of $\chi$ as  
    \begin{equation}
      \Val_{\chi}(\mathcal{F}) = \sum\limits_{r=1}^m w_r P_r(\chi(S_r)).
    \end{equation}
    We also define the optimal value of $\mathcal{F}$ in the case of Max-CSP to be 
    \begin{equation}
      \Opt(\mathcal{F}) = \max\limits_{\chi} \left( \Val_{\chi}(\mathcal{F}) \right).
    \end{equation}
    In the minimization version, correct definition of the optimal value has ``$\min$'' instead of ``$\max$'' in the previous expression. Typically, the aim is to find a solution with the  value close to the optimal one. In case of Max-CSP, an $\alpha$ approximation algorithm is an algorithm which in polynomial time finds an assignment $\chi$ such that


    \begin{equation*}
      \Val_{\chi}(\mathcal{F}) \geq \alpha \cdot \Opt(\mathcal{F}). 
    \end{equation*}
    For Min-CSPs, the correct definition has ``$\leq$'' instead of ``$\geq$'' in the previous inequality.
    \par 
		While introducing weights allows convenient representation of CSPs, the hardness of approximation \emph{essentially} does not change, as shown in \cite[Lemma 3.11]{DBLP:journals/siamcomp/KhannaSTW00}. We reprove these results here, starting with a following lemma.


    \begin{lemma} \label{weighted_vs_unweighted_lemma_deterministic}
      Consider a weighted instance $\mathcal{F}$ of a Max-CSP (or Min-CSP) $\Lambda$. Then, for each $\varepsilon>0$ there is a poly-time algorithm which outputs an unweighted instance $\mathcal{G}$ of the same CSP $\Lambda$ over the same variables as in $\mathcal{F}$ such that 
      \begin{equation}
        \Val_{\zeta}(\mathcal{G}) -\varepsilon \leq \Val_{\zeta}(\mathcal{F}) \leq \Val_{\zeta}(\mathcal{G})  + \varepsilon,
      \end{equation}
      where $\zeta$ is any assignment to variables of $\mathcal{F}$ (or $\mathcal{G}$). Furthermore, the size of instance $\mathcal{G}$ is polynomial in size of $\mathcal{F}$ and $1/\varepsilon$.
      \begin{proof}
        Let $\mathcal{F}$ be an instance over constraints $C_1,\hdots,C_m$ with respective weights $w_1,\hdots,w_m$. We fix $q = \lceil m / \varepsilon \rceil$, and construct $\mathcal{G}$ by creating $\ell_r$ copies of each constraint $C_r$, where $\ell_r$ are chosen such that
        \begin{equation}
          \begin{split}
						\sum_r \ell_r = q, \quad \frac{\ell_r}{q} \in \left(w_r-\frac{1}{q},w_r+\frac{1}{q}\right).
          \end{split}
        \end{equation}
				We can find such $\{\ell_r\}_{r=1}^m$ by setting first $\ell_r = \lfloor w_r q \rfloor $, and then incrementing some $\ell_i$  to obtain $\sum_r \ell_r = q$. \par 
        For any given assignment $\zeta$ to the variables, the contribution towards the value of $\mathcal{F}$ of each constraint $C_r$ in $\mathcal{F}$ is at most $1/q$ different from contributions of replicated constraints in $\mathcal{G}$. Finally, since we have $m$ constraints, the main claim
      \begin{equation}
        \Val_{\zeta}(\mathcal{G}) -\varepsilon \leq \Val_{\zeta}(\mathcal{F}) \leq \Val_{\zeta}(\mathcal{G})  + \varepsilon
      \end{equation}
      of the theorem follows.
      \end{proof}
    \end{lemma}
		By relying on this lemma, we can show that weights do not affect approximability of Max/Min-CSPs, as long as we allow for additive $\varepsilon$ loss in approximation ratio. We first prove this claim for Max-CSPs. 
    \begin{theorem}\label{max_csp_weighted_theorem}
      Consider a Max-CSP $\Lambda$ and assume that we can approximate the optimal value of unweighted instances within a multiplicative factor $\alpha$. Then, for every $\delta>0$, weighted instances of Max-CSP $\Lambda$ can be approximated within a constant $\alpha-\delta$.
      \begin{proof}
				{\color{NavyBlue}
			Without loss of generality let us assume that $\Lambda$ does not contain a predicate $P \equiv 0$, since we can remove each constraint with a predicate $P \equiv 0$ from an instance and rescale the weights, which does not affect approximability in the discussion that follows since the ratio between the values under any two assignments remains the same.  \par
        Now, let us fix a weighted instance $\mathcal{F}$ of a CSP $\Lambda$, and consider a random assignment $\chi$ in which each variable takes value $0$ or $1$ with probability $1/2$, independently. Then, the expected value of $\mathcal{F}$ under this random assignment is 
        \begin{equation*}
          \E_{\chi}\left[ \sum_{r=1}^m w_r P_r(\chi(S_r))	 \right] = \sum_{r=1}^m w_r \E_{\chi}\left[ P_r(\chi(S_r))\right]. 
        \end{equation*}
			The value $\E_{\chi}[P_r(\chi(S_r)) ]$ depends only on the properties of the predicate $P_r$. Furthermore, by our assumption $P_r \not \equiv 0$, and therefore $\E_{\chi}[P_r(\chi(S_r))]>0$. Thus, since $\{P_r\}_{r=1}^m$ are picked from a finite collection of predicates of $\Lambda$, there is a $\gamma>0$ such that $\E_{\chi}[P_r(\chi(S_r))] \geq \gamma$, for every $r \in [m]$. Therefore, we have that  }
        \begin{equation*}
          \E_{\chi}\left[ \sum_{r=1}^m w_r P_r(\chi(S_r))	 \right] = \sum_{r=1}^m w_r \E\left[ P_r(\chi(S_r))\right] \geq \sum_{r=1}^m w_r \gamma =\gamma  .
        \end{equation*}
        Hence, under the randomized assignment, the instance has a value of at least $\gamma $ in expectation. By the averaging argument, we have that $\Opt(\mathcal{F})\geq \gamma$. Now, consider the algorithm from Lemma \ref{weighted_vs_unweighted_lemma_deterministic} with parameter $\varepsilon = \delta \gamma /2$, which takes our instance $\mathcal{F}$ and outputs unweighted instance $\mathcal{G}$. We can apply the $\alpha$ approximation algorithm on $\mathcal{G}$ to obtain some assignment $\zeta$ for which
        \begin{equation*}
          \frac{\Val_{\zeta}(\mathcal{G})}{\Opt(\mathcal{G})} \geq \alpha.
        \end{equation*}
				Then, since $\Opt(\mathcal{G}) \geq \gamma $, for the same assignment $\zeta$ we have 
        \begin{equation*}
          \begin{split}
            \frac{\Val_{\zeta}(\mathcal{F})}{\Opt(\mathcal{F})} \geq \frac{\Val_{\zeta}(\mathcal{G}) -\varepsilon}{\Opt(\mathcal{G})+\varepsilon}   \geq  
            \frac{\Val_{\zeta}(\mathcal{G})-\varepsilon}{\Opt(\mathcal{G})}   ( 1 - \varepsilon/\gamma) \geq  \\ 
            \frac{\Val_{\zeta}(\mathcal{G})}{\Opt(\mathcal{G})} -\frac{\varepsilon}{\Opt(\mathcal{G})}  - \frac{\varepsilon \Val_{\zeta}(\mathcal{G})}{\gamma\Opt(\mathcal{G})} + \frac{\varepsilon^2}{\gamma \Opt(\mathcal{G})}
            \geq \alpha - \frac{\varepsilon}{\gamma}- \frac{\varepsilon}{\gamma}  \geq \alpha-\delta ,
          \end{split}
        \end{equation*}
        which proves the statement of the theorem.
      \end{proof}
    \end{theorem}
		The argument from the previous theorem does not work for Min-CSPs, since in this case $\Opt(\mathcal{G})$ can be arbitrarily small. Analogous claim for Min-CSPs was already proved in \cite[Lemma 3.11]{DBLP:journals/siamcomp/KhannaSTW00} by using scaling techniques \cite{DBLP:journals/jacm/IbarraK75,DBLP:conf/istcs/CrescenziST96}. For the sake of completeness, we give here somewhat more detailed proof of this claim, using essentially the same techniques. 
		\begin{theorem} \label{min_csp_det_theorem}
      Consider a Min-CSP $\Lambda$, and assume we can approximate the optimal value of unweighted instances within a multiplicative factor $\alpha$. Then, for every $\delta \in (0,1)$, weighted instances of the Min-CSP $\Lambda$ can be approximated within a constant $\alpha+\delta$.
      \begin{proof}
        Consider the decision version of CSP $\Lambda$, in which we ask whether there is an assignment $\chi$ to the variables such that all the constraints of $\Lambda$ are \emph{not} satisfied. By Schaefer's dichotomy theorem \cite{Schaefer:1978:CSP:800133.804350}, the problem of deciding whether there is an assignment which falsifies all the constraints is either $\NP$-hard or in $\Pal$. If solving this problem is $\NP$-hard, then both weighted and unweighted versions of Min-CSP $\Lambda$ are obviously $\NP$-hard to approximate within any constant. Therefore, without loss of generality, we assume that deciding whether all constraints can be falsified is in $\Pal$ for $\Lambda$.
        \par 
      Hence, given an instance $\mathcal{F}$ of the Min-CSP $\Lambda$, we can check in polynomial time whether $\Opt(\mathcal{F})=0$. In case $\Opt(\mathcal{F})=0$, we have found an optimal assignment, so it only remains to consider $\Opt(\mathcal{F})>0$.
      \par Without loss of generality let us assume that the weights of constraints $\{w_i\}_{i=1}^m$ are sorted in descending order, i.e. $w_1 \geq w_2 \geq \hdots \geq w_m$. We can find in polynomial time the largest $k \geq 1$ such that there is an assignment falsifying constraints $C_{1},C_{2},\hdots,C_{k-1}$.  \par 
				For thusly chosen $k$ at least one of $C_1,\hdots,C_{k}$ will be true in any assignment, so we have that $\Opt(\mathcal{F}) \geq w_{k}$. Also, since there is an assignment falsifying the first $k-1$ constraints, we have that $\Opt(\mathcal{F}) \leq \sum_{i=k}^{m} w_i$. 
        \par 
        Let us partition the constraints $C_i$ into the following three groups:
        \begin{itemize}
          \item \emph{light}: constraints $C_i$ with weight $w_i \leq w_k/m^2$. 
          \item \emph{medium}: constraints $C_i$ with weight $w_k/m^2 < w_i < w_k m^2 $. 
          \item \emph{heavy}: constraints $C_i$ with weight $w_k m^2 \leq w_i $.
        \end{itemize}
        Then, we create an instance $\mathcal{F}'$ by adding medium and heavy constraints $C_i$ from $\mathcal{F}$. Furthermore, we scale down the weights of heavy constraints to $w_k m^2 $ in $\mathcal{F}'$.  Finally, we normalize the weights to total weight by multiplying them by some factor $\sigma >1$. Note that $\Opt(\mathcal{F'}) \geq w_k \sigma $, since $\mathcal{F}'$ still has (although with different weights) constraints $C_1,C_2,\hdots,C_k$. Hence, in a completely analogous manner to Theorem \ref{max_csp_weighted_theorem}, we can use the algorithm from Lemma \ref{weighted_vs_unweighted_lemma_deterministic} with $\varepsilon = \frac{\delta w_k \sigma}{4\alpha} $, to get an assignment $\zeta$ which gives us an $\alpha+\delta/2$ approximation of the optimal value for $\mathcal{F}'$. Finding the $\alpha+\delta/2$ approximation can be performed in polynomial time, because $w_k \sigma \geq m^{-3}$,
        \par 
        Let us now see how well $\zeta$ approximates the optimal value of $\mathcal{F}$. We have that following two properties:
        \begin{itemize}
          \item \textbf{Property A:}  $\Opt(\mathcal{F}) \geq \frac{1}{\sigma}\Opt(\mathcal{F}')$. This holds since optimal values of both $\mathcal{F}$ and $\mathcal{F}'$ do not satisfy heavy constraints. 
        \item \textbf{Property B:} If an assignment $\chi$ does not satisfy heavy constraints, then $\Val_{\chi}(\mathcal{F}) \leq \frac{1}{\sigma}\Val_{\chi}(\mathcal{F}')+w_k/m $ . This statement holds since if we do not satisfy heavy (scaled down) constraints, then the only difference comes from light constraints, which can have a total weight of at most $w_k/m$.   
      \end{itemize}
      Finally, consider our $(\alpha+\delta/2)$-approximating assignment $\zeta$ to the instance $\mathcal{F}'$. This assignment certainly does not satisfy heavy constraints, since otherwise we have $\Val_{\zeta}(\mathcal{F}') \geq w_k m^2 \sigma$, and $\Opt(\mathcal{F}')\leq w_k m  \sigma$, so the approximation ratio would be at least $m$, which can not happen since $\zeta$ achieves a constant factor approximation. Therefore, by using properties \textbf{A} and \textbf{B} for $\zeta$ we have 
        \begin{equation*}
          \begin{split}
            \frac{\Val_{\zeta}(\mathcal{F})}{\Opt(\mathcal{F})} = \frac{\sigma\Val_{\zeta}( \mathcal{F})}{\sigma \Opt(\mathcal{F})} \leq \frac{\Val_{\zeta}(\mathcal{F}')+\sigma w_k/m}{\Opt(\mathcal{F}')} \leq \frac{\Val_{\zeta}(\mathcal{F}')}{\Opt(\mathcal{F}')} + \frac{\sigma w_k/m}{\sigma w_k} \leq \alpha +\delta/2 + \frac{1}{m}.
          \end{split}
        \end{equation*} 
				Since $1/m < \delta/2$ for sufficiently $m$, our theorem holds.
      \end{proof}
    \end{theorem}


  \section{Reduction}\label{red_section}
	We now prove the theorem which shows the existence of a randomized algorithm which can be used for proving Theorem \ref{main_theorem_introduction_max_random}. We remark that this theorem uses a reduction that appeared in \cite{DBLP:conf/stoc/Trevisan01}, and that the main difference comes from the fact that we need to create instances in which degrees of variables are uniform, while bounded degree was sufficient in \cite{DBLP:conf/stoc/Trevisan01}. Additional complexity lies in the fact that we prove theorem for any Max-CSP, and therefore account for different arity of predicates, while \cite{DBLP:conf/stoc/Trevisan01} considered Max-$E3$-Sat with predicates of arity $3$.
	\par 
	Let us now give an overview of the proof. We start in the same way as \cite{DBLP:conf/stoc/Trevisan01}, by creating $d_i$ \emph{copies} $x_i^j, j =1,\hdots,d_i$ for each variable $x_i$ in the starting instance of degree $d_i$. Then, in order to create a regular instance, we sample constraints of the starting instance, and create a constraint in the new instance by replacing each $x_i$ occurring in the scope by some of its copies $x_i^j$ uniformly at random. Such a procedure outputs an instance in which every variable has the same degree in expectation. However, with small probability it can still happen that the deviations from this degree are large. For that reason, we repeat this procedure up to $n$ times until the degrees of the variables are close to the expected value, or otherwise our algorithm fails. In case of our algorithm not failing, we slightly update the resulting instance to ensure that each variable has the same degree. More precisely, in case expected degree of each variable is $D$, we replace variables with degree higher than $(1+\beta)D$ in scopes of some constraints with some new \emph{dummy} variables, where $\beta$ is small. Finally, we also create new constraints in order to make sure that each variable $x_i^j$ has degree exactly $(1+\beta)D$. Final step in our construction consists in making sure that newly introduced \emph{dummy} variables also have degree $(1+\beta)D$. We then show that with very high probability these updates changed/added only small number of constraints, so our regular instance "looks like" the random one. The last part of the proof shows that an assignment to regular instance can be used to construct an assignment which satisfies similar fraction of constraints of the starting instance. The idea is the same as in \cite{DBLP:conf/stoc/Trevisan01}; namely, the fraction of variables $x_i^j$ with value $1$ gives us the probability that variable $x_i$ should have value $1$, and this is used in randomized algorithm which converts the values of $x_i^j$ to values of $x_i$. This algorithm can be derandomized, and we show that this conversion does not incur large change in the value of the instance.
	\par 
		A formal statement and a proof are given below.
  \begin{theorem} \label{max_csp_rand_theorem}
		Consider an unweighted instance $\mathcal{F}$ of a Max-CSP $\Lambda$ and let  $\varepsilon>0$. Then, there is a randomized algorithm which outputs a regular instance $\mathcal{G}$ of the Max-CSP $\Lambda$ such that  with probability at least $1-O(2^{-n})$ over the choices made in the randomized algorithm, the following two statements hold:
    \begin{itemize}
      \item[\textbf{[a]}]  For any assignment $\zeta$ to the variables of $\mathcal{G}$, there is an algorithm which runs in polynomial time and finds an assignment $\chi$ to the variables of $\mathcal{F}$ such that
        \begin{equation*}
					\Val_{\chi}(\mathcal{F}) \geq \Val_{\zeta}(\mathcal{G}) -\varepsilon.
        \end{equation*}
      \item[\textbf{[b]}]  The optimal value of $\mathcal{F}$ is upper bounded by $\Opt(\mathcal{G}) + \varepsilon$.
    \end{itemize}
		Furthermore, the runtime of the randomized algorithm is polynomial in terms of the size of $\mathcal{F}$ and $\lceil 1/\varepsilon \rceil$, and the degree of variables in $\mathcal{G}$ is $O(-\log(\varepsilon)/\varepsilon^2)$.
    \begin{proof}
      We begin by describing a randomized procedure that creates the regular instance $\mathcal{G}$. The instance $\mathcal{F}$ contains $m$ constraints $C_1,\hdots,C_m$, and each constraint is applied to some tuple of variables $S_i$. First, for each variable $x_i$ from $\mathcal{F}$ with degree $d_i$, we create $d_i$ new variables $x_i^1,\hdots,x_i^{d_i}$.  Then we fix $D \in \mathbb{N}$, which will be suitably chosen later,  and create an instance $\mathcal{G}'$ with $mD$ constraints over variables $x_i^j$ by repeating the following procedure $mD$ times:
    \begin{itemize}
      \item Pick a constraint $C_i=(P_i,S_i)$ from $\mathcal{F}$ uniformly at random.
      \item For each each variable $x_j$ appearing in $S_i$, pick a variable $x_j^r$ from a set $\{x_j^1,\hdots,x_j^{d_j}\}$ uniformly at random.
      \item Add a constraint $C_i'=(P_i,S_i')$ to $\mathcal{G}'$, which constrains the variables $x_j^r$ picked in the previous step by the predicate $P_i$ of the constraint $C_i$. Furthermore, each variable $x_j^r$ appears at the same position in the tuple $S_i'$ as the variable $x_j$ in the tuple $S_i$.
    \end{itemize}
    Since each variable $x_i^j$ is picked at each step with probability $1/m$, every variable $x_i^j$ in $\mathcal{G}'$ in expectation has degree $D$. However, some variables can have larger (or smaller) degree than $D$. For that reason, we will update the instance $\mathcal{G}'$ by changing and adding some constraints to create an instance $\mathcal{G}$ in which each variable has degree $(1+\beta)D$, where $\beta>0$ is a number that will be fixed later. By Chernoff's bound (Lemma \ref{Chernoff2}), the probability that a variable $x_i^j$ appears in more than $(1+\beta)D$ constraints in $\mathcal{G}'$ is given by
    \begin{equation}
      \Prob \left[  \deg(x_i^j)   \geq (1+\beta ) D \right] \leq 2 e^{-\frac{\beta ^2 D}{4} }.
    \end{equation}
    If variable $x_i^j$ appears $t>(1+\beta)D$ times, we replace $x_i^j$ by some new variable in $t-(1+\beta)D$ constraints. The expected number of times we need to do this for a fixed $x_i^j$ is at most 
		\begin{equation*}
			\begin{split}
				\sum_{t=(1+\beta)D+1}^{mD} \Pr \left[ \deg(x_i^j) \geq t \right]  (t-(1+\beta)D) \leq 		2 \sum_{t=(1+\beta)D}^{\infty} \exp\left( -\frac{\left(\frac{t}{D}-1\right)^2}{4} D   \right)(t-(1+\beta) D) \\
				\leq 2 \sum_{t=(1+\beta)D}^{\infty} \exp \left( -\frac{(t-D)^2}{4D}  \right)  (t-(1+\beta) D)
			\end{split}
		\end{equation*}
		By introducing $c=t-(1+\beta)D$, we can further simplify this expression as  
		\begin{equation*}
			\begin{split}
        2 \sum_{c=0}^{\infty} \exp\left(-\frac{(\beta D + c)^2}{4 D }\right) c & \leq 2 \sum_{c=0}^{\infty} \exp\left(-\frac{1}{4}\frac{(\beta D + c)^2}{\frac{\beta}{\beta} D + \frac{1}{\beta} c}\right) c \\
				\leq 2 \sum_{c=0}^{\infty} \exp\left(-\frac{\beta}{4}(\beta D + c) \right) c  & \leq 2 \int_{0}^{\infty} \exp\left(-\frac{\beta}{4}(\beta D + c) \right) c \diff c
			\end{split}
		\end{equation*}
    We can show that this expression is smaller than $1$ for $D = \Omega\left(-\log(\beta)/\beta^2\right)$. Therefore, for such $D$, we replace each variable $x_i^j$ at most once in expectation. Finally, by Markov's inequality we conclude that with probability at least $3/4$ we need to replace at most $4mW$ variables in $\mathcal{G}'$, where $W$ is the average arity of constraints in $\mathcal{F}$. If we need to replace more than $4mW$ variables, the construction of $\mathcal{G}'$ fails. Otherwise, the degree of each variable $x_i^j$ is at most $(1+\beta)D$.
		\par
    Let us now consider variables $x_i^j$ such that  $\deg(x_i^j) < (1+\beta)D$. For each such variable $x_i^j$ we create $(1+\beta)D - \deg(x_i^j)$ new constraints, in order to make sure that each variable $x_i^j$ has degree $(1+\beta)D$. Note that in this process we might need to add some additional \emph{dummy} variables, which can have degree different from $(1+\beta)D$. We will handle these variables later. Let us now estimate how many constraints we need to add. In order to do that, we first find an upper bound on the probability that the average arity of constraints in $\mathcal{G}'$ is smaller than $(1-\beta)W$. In particular, if we denote with $W_{\max}$ the maximal arity of the constraints in $\mathcal{F}$, application of Hoeffding's  inequality (Lemma \ref{hoeff_bound}) gives
    \begin{equation*}
      \Pr \left[ \left| \frac{1}{mD} \sum_{C_i' \in \mathcal{G}'}\ar(P_i') - W \right |   \geq \beta W  \right] \leq  2 \exp\left(-\frac{\beta^2 m^2D^2W^2 }{m D W_{\max}^2}\right) = 2 \exp\left(-\frac{\beta^2 mDW^2 }{W_{\max}^2}\right).
		\end{equation*}
		Observe that for $D=\Omega\left(\frac{W_{\max}^2}{\beta^2 m}\right)$ this value is smaller than $1/4$. If average arity of constraints in $\mathcal{G}'$ is smaller than $(1-\beta)W$, construction of $\mathcal{G}'$ fails. Otherwise, we add up to $2\beta m W D + 4mW$ constraints. After this stage of the algorithm, each variable $x_i^j$ will have degree exactly $(1+\beta)D$. 
		\par 
		By union bound the probability that the construction of $\mathcal{G}'$ fails is at most $1/2$. We actually try to construct $\mathcal{G}'$ up to $n$ times, and therefore the probability of failure for constructing $\mathcal{G}'$ drops to $2^{-n}$. In case all $n$ trials of constructing $\mathcal{G}'$ are unsuccessful, we stop the execution of the algorithm and report failure. Otherwise, we proceed in the manner described below. 
    \par 
		In the process of ensuring that the degree of each variable $x_i^j$ is exactly $(1+\beta)D$, we introduced some additional dummy variables. At each step of updating $\mathcal{G}'$ we only need to keep $W_{\max}-1$ additional dummy variables that were used less than $(1+\beta)D$ times, because the {\color{NavyBlue}largest arity of constraints in $\mathcal{F}$ is $W_{\max}$}. Therefore, we can assume that after this process we are left with at most $W_{\max}$ dummy variables with degree smaller than $(1+\beta)D$, and therefore to ensure they have degree $(1+\beta)D$ we need to use them $t < (1+\beta)D W_{\max} $ times. If we pick $\beta$ such that $(1+\beta)D$ is coprime\footnote{We can do this for all sufficiently small $\varepsilon$, by our choice of $D$ which we discuss later.} with $W_{\max}$, then  there is some $k$ such that $W_{\max}\leq k < 2W_{max}$ and $W_{max}  \mid k(1+\beta)D + t$. Let us then introduce $k$ new dummy variables and add $(k(1+\beta)D+t)/W_{\max}$ constraints with predicate of arity $W_{\max}$, by assigning the scopes such that the degree of each variable becomes exactly $(1+\beta)D$. In particular, we can assign scopes iteratively, by adding variables with the smallest degree to the scope at each step. Since $k\geq W_{\max}$ and $W_{\max}$ variables have the degree $0$ at the start, we can always assign $W_{\max}$ distinct variables to the scope. Finally, since $W_{\max} \mid k(1+\beta)D +t$ the iterative procedure will finish with all variables having degree $(1+\beta)D$ after exactly $(k(1+\beta)D+t)/W_{\max}$ steps.
		\par Summarily, we have introduced or changed at most the following number of constraints:
		\begin{itemize}
      \item $4 mW$ constraints by replacing variables $x_i^j$ by new dummy variables, in order to make sure that no variable $x_i^j$ has degree bigger than  $(1+\beta)D$.
			\item $2 \beta m W D + 4mW$ constraints in order to make sure that every variable $x_i^j$ does not have degree smaller than $(1+\beta )D$.
      \item $3 (1+\beta) D $ new constraints introduced to ensure that additionally added variables have degree $(1+\beta)D$. 
		\end{itemize}
		In particular, the number of constraints that we did not change is at least $mD-4mW$, and the number of new or changed constraints is at most 
		\begin{equation*}
      B := 8mW + 2 \beta m W D + 3(1+\beta)D .
		\end{equation*}
    This concludes the description of $\mathcal{G}$. Let us now prove statement \textbf{[a]}. \par 
    Let $\zeta$ be an assignment to variables of $\mathcal{G}'$. Then, consider a randomized assignment $\bar{\chi}$ to the variables of $\mathcal{F}$, in which variables $x_i$ get values independently, and the probability of $x_i$ getting the value $1$ is proportional to the number of variables $x_i^j$ getting the value $1$ under $\zeta$. Let us denote with $\rho_{\zeta}$ the expected value of $\mathcal{F}$ under random $\bar{\chi}$. 
		\par 
    Consider now one trial in the process of creating $\mathcal{G}'$ from $\mathcal{F}$. At each step, we pick a constraint $C_i$, which is satisfied by $\bar{\chi}$ with probability $\rho_{\zeta}$. Furthermore, the respective constraint $C_i'$ in $\mathcal{G}'$ is satisfied by $\zeta$ with probability $\rho_{\zeta}$ as well. Therefore, our algorithm will create an instance $\mathcal{G}'$ that under $\zeta$ satisfies a fraction $\rho_{\zeta}$ of constraints in expectation. By Chernoff's bound (Corollary \ref{Chernoff1}), the probability that the fraction of satisfied constraints in $\mathcal{G}'$ is bigger than $\rho_{\zeta} + \varepsilon/2$ is at most $2 \exp\left( -\varepsilon^2 m D /16 \right)$.
    Therefore, if we pick $D=\Omega(\frac{W}{ \varepsilon^2})$, we have that 
    \begin{equation*}
      \Pr\left[ \Val_{\zeta}(\mathcal{G}') \geq \rho_{\zeta}+ \varepsilon/2 \right] \leq 2^{ -2mW }.
    \end{equation*} 
		Since there are $mW$ variables $x_i^j$ in $\mathcal{G}'$, there are $2^{mW}$  possible assignments $\zeta$. Hence, by union bound the probability that there is an assignment $\zeta$ which satisfies more than $\rho_{\zeta} + \varepsilon/2$ constraints in $\mathcal{G}'$ is at most $2^{-mW}$. Furthermore, since a trial succeeds with probability at least $1/2$, the probability that no assignment satisfies more than $\rho_{\zeta}+\varepsilon/2$ constraints in $\mathcal{G}'$ conditioned on the fact that the trial is successful is at least $1-2 \cdot 2^{-mW}$.\par
		Finally, when switching from $\mathcal{G}'$ to $\mathcal{G}$, in the worst case we can have at most $B$ more satisfied constraints, and hence
		\begin{equation*}
      \Val_{\bar{\chi}}(\mathcal{F})+\varepsilon/2 \geq \Val_{\zeta}(\mathcal{G}') \geq \Val_{\zeta}(\mathcal{G}) - \frac{B}{mD}. 
		\end{equation*}
		By choosing $\beta = \frac{\varepsilon}{5W}$, and for sufficiently big $m$ and $D$ we have that $\frac{B}{mD} \leq \varepsilon /2 $, and therefore we have that 
		\begin{equation*}
     \Val_{\bar{\chi}}(\mathcal{F}) \geq \Val_{\zeta}(\mathcal{G})  - \varepsilon.
		\end{equation*}
		This inequality holds for the randomized assignment $\bar{\chi}$. However, using the method of conditional probabilities we can efficiently and deterministicaly find an assignment $\chi$ such that $	\Val_{\chi}(\mathcal{F}) \geq \Val_{\bar{\chi}}(\mathcal{F})$. This proves that \textbf{[a]} holds with probability at least $1- 2\cdot 2^{-mW}-2^{-n}$ over the random choices in the algorithm that creates $\mathcal{G}$.
		\par 
		The statement \textbf{[b]} can be proved in a similar way. In particular, let us fix the assignment $\chi$ to the variables of $\mathcal{F}$ under which the optimal value $c:=\Opt(\mathcal{F})$ is attained. Then, we construct an assignment $\zeta$ for $\mathcal{G}$, by setting $x_i^j$ to have the same values as the corresponding $x_i$ under $\chi$, and we assign the values to the remaining variables arbitrarily. Throughout randomized construction of $\mathcal{G}'$ in one trial, each constraint is satisfied with probability $c$, and therefore the expected fraction of satisfied constraints in $\mathcal{G}'$ is $c$. The probability (over the random choices made in a trial) that $\Val_{\zeta}(\mathcal{G}') \geq c+\varepsilon/2$ is by Chernoff's bound (Corollary \ref{Chernoff1}) at most 
    \begin{equation*}
      2 \exp\left(\frac{-mD\varepsilon^2}{16} \right).
    \end{equation*}
		Therefore, for $D=\Omega\left(\frac{W_{\max}}{\varepsilon^2}\right)$, this probability will be smaller than $2^{-n}$. Moreover, by noting that changing or adding $B$ constraints when switching from $\mathcal{G}'$ to $\mathcal{G}$ can not impact the optimal value by more than $\varepsilon/2$, we see that the statement $\textbf{[b]}$ holds with probability $1-3\cdot 2^{-n}$, over the random choices in the algorithm that creates $\mathcal{G}$. By union bound the statements \textbf{[a]} and \textbf{[b]} both hold with probability at least $1-2\cdot2^{-mW}-4\cdot 2^{-n}=1-O(2^{-n})$, which concludes the proof of this theorem.
 		\end{proof} 
	\end{theorem}
  Let us now prove Theorem \ref{main_theorem_introduction_max}. For that reason, let us suppose that regular instances of Max-CSP $\Lambda$ can be approximated within some fixed approximation ratio $\alpha$, and let us consider an arbitrary (possibly not regular) instance $\mathcal{F}$ of $\Lambda$. Then, we construct $\mathcal{G}$ with the probabilistic algorithm described in the previous theorem, apply the $\alpha$ approximation algorithm to find an assignment $\zeta$, and from  $\zeta$ using the algorithm from Theorem \ref{max_csp_rand_theorem} \textbf{[a]}, we can find an assignment $\chi$ to the instance $\mathcal{F}$ satisfying
  \begin{equation*}
    \begin{split}
      \frac{\Val_{\chi}(\mathcal{F})}{\Opt(\mathcal{F})} \geq \frac{\Val_{\chi}(\mathcal{G})-\varepsilon}{\Opt(\mathcal{G})+\varepsilon} & \geq 
    \frac{\Val_{\zeta}(\mathcal{G})-\varepsilon}{\Opt(\mathcal{G})} \left(1-\frac{\varepsilon}{\Opt(\mathcal{G})}\right) \\
                                                                                                                                         & \geq  \frac{\Val_{\zeta}(\mathcal{G})}{\Opt(\mathcal{G})} - \frac{\varepsilon \Val_{\zeta}(\mathcal{G})}{\Opt(\mathcal{G})^2} - \frac{\varepsilon}{\Opt(\mathcal{G})} \geq \alpha - 2\frac{\varepsilon}{\Opt(\mathcal{G})} . 
    \end{split}
  \end{equation*}
  Now, we have that $\Opt(\mathcal{G})\geq \gamma$, for some\footnote{As in the proof of Theorem \ref{max_csp_weighted_theorem}, w.l.o.g.~we assume that instance $\mathcal{F}$ does not contain predicates which evaluate to $0$ under all assignments.}  fixed $\gamma>0$ which depends only on $\Lambda$. Therefore, by choosing $\varepsilon= \delta \gamma /2 $, the claim of Theorem  \ref{main_theorem_introduction_max} follows.
   \par
	 Note that using analog of Theorem \ref{max_csp_rand_theorem} for Min-CSPs to prove Theorem \ref{main_theorem_introduction_min} would require $\varepsilon=O(1/m)$, and therefore the instance $\mathcal{G}$ in the reduction will be of size at least $m^3\log(m)$, with $D=\Omega(m^2\log(m))$. We give a deterministic reduction instead, which works for both Max-CSPs and Min-CSPs, and which creates regular instance of degree $\lceil \bar{D}/\varepsilon \rceil$  where $\bar{D}$ is the maximal degree of constraints in $\mathcal{F}$.  The reduction is given as the following theorem.
  \begin{theorem} \label{max_min_csp_det_theorem}
		Consider a Max-CSP (or Min-CSP) $\Lambda$ and let  $\varepsilon>0$.  Then, there is a reduction which takes as an input an instance $\mathcal{F}$ of a Max-CSP (or Min-CSP)$\mathcal{F}$ and outputs a regular instance $\mathcal{G}$ of the Max-CSP (or Min-CSP) $\Lambda$ such that the following holds
    \begin{itemize}
      \item[\textbf{[a]}]  $\Opt(\mathcal{F})  \leq \Opt(\mathcal{G}) $ (or $\Opt(\mathcal{F}) \geq \Opt(\mathcal{G})$ for Min-CSP)
      \item[\textbf{[b]}]  Let $\zeta$ be an assignment to the variables of $\mathcal{G}$. There, there is an algorithm which finds an assignment $\chi$ to the variables of $\mathcal{F}$ such that
        \begin{equation*}
          \begin{split}
            \Val_{\chi}(\mathcal{F}) \geq \Val_{\zeta}(\mathcal{G}) -\varepsilon  \qquad \qquad \textrm{\emph{(Max-CSP case)}},\\
            \Val_{\chi}(\mathcal{F}) \leq \Val_{\zeta}(\mathcal{G}) +\varepsilon \qquad \qquad \textrm{\emph{(Min-CSP case)}}.
          \end{split}
        \end{equation*}
				Furthermore, the runtime of the reduction and of the algorithm from \textbf{[b]} is polynomial in terms of the size of $\mathcal{F}$ and $\lceil 1/\varepsilon \rceil$.
    \end{itemize}
    \begin{proof}
      We prove this theorem for Min-CSP $\Lambda$. The proof for Max-CSP is analogous.
      \par
      We begin by describing how a regular instance $\mathcal{G}$ is constructed. We start from $\mathcal{F}$ which has $n$ variables $x_1,\hdots,x_n$, with degrees $d_1,\hdots,d_n$, and constraints $\{(P_r,S_r)\}_{r=1}^m$. For each variable $x_i$ from $\mathcal{F}$ we create $d_i$ variables $x_i^1,x_i^2,\hdots,x_i^{d_i}$ in $\mathcal{G}$. The constraints of $\mathcal{G}$ are constructed as follows. We begin by creating $N$ copies of $\mathcal{F}$ which we call blocks. Then, we go through the blocks and at each scope we replace variables $x_i$ by their corresponding copies $x_i^j$, $j \in [d_i]$. In particular, each $x_i$ can be replaced only by $d_i$ variables $x_i^1,x_i^2,\hdots,x_i^{d_i}$. Since $x_i$ appears $d_i$ times in $\mathcal{F}$, it will get replaced $N d_i$ times, and in order to impose regularity we replace $x_i$ by each copy $x_i^j$ exactly $N$ times. Therefore, the degree of all variables is $N$, and instance $\mathcal{G}$ is regular.
      \par 
      Actually, we will be a bit more careful when replacing variables $x_i$ by their copies $x_i^j$. The idea is that each block should resemble $\mathcal{F}$ as much as possible, and therefore we want to avoid replacing $x_i$ by two different copies $x_i^j, x_i^k$,  in the same block. Let us call a block \emph{good} if each variables $x_i$ is replaced by a single copy $x_i^j$ in that block. Our aim is to maximize the number of good blocks, which we do greedily by creating a good block at each step of the algorithm as long as we can, after which each variable $x_i$ is replaced by any of its available copies. Let us remark that at all stages of our algorithm we still make sure that each copy is not used more than $N$ times, and that each $x_i$ is replaced only by its own copies. \par 
      This finishes our description of $\mathcal{G}$. Before we prove the claims of the theorem, let us find a lower bound on the number of good blocks created. In our greedy algorithm, a good block can be created if for each variable $x_i$ we can find $x_i^j$ which was used $N-d_i$ times or less. Therefore, each variable $x_i^j$ can be used at least $\lfloor N/d_i \rfloor$ times for creating a good block, and therefore the number of good blocks is $\min_{i \in [n]} \lfloor N / d_i \rfloor d_i$. Hence, by letting $D= \max_{i \in [n]}d_i$, we conclude that there are at least $N-D$ good blocks. 
    \par 
    We construct $\mathcal{G}$ as described above with $N=\lceil D/\varepsilon \rceil$. It is straightforward to verify \textbf{[a]}, and hence let us now prove the claim \textbf{[b]}. For a given assignment $\zeta$ of variables in $\mathcal{G}$, let us consider a good block with the smallest value, and let us denote the value of this block by $v$. We define an assignment $\chi$ to be the assignment of $\zeta$ on this block. We note that we can do this since the copies of $x_i$ are unique in every good block. We have that $\Val_{\chi}(\mathcal{F}) = v$, and therefore since $v$ is the minimal value of good blocks we have 
    \begin{equation*}
      \Val_{\zeta}(\mathcal{G}) \geq \frac{1}{N}\left( (N-D) v \right )  \geq v - \frac{D}{N} = \Val_{\chi}(\mathcal{F}) - \varepsilon,
    \end{equation*}
    which finishes the proof of \textbf{[b]}. 
    \end{proof}
  \end{theorem}
  Let us now show how this result can be used to prove Theorem \ref{main_theorem_introduction_min}. Hence, let us fix $0<\delta<1$, and starting from an instance $\mathcal{F}$ of a Min-CSP $\Lambda$ with constraints $C_1,\hdots,C_m$, we apply algorithm from the previous theorem with $\varepsilon = \delta/m$ to get a regular instance $\mathcal{G}$. Then, we use the $\alpha$ approximation algorithm to get an assignment $\zeta$ to variables of $\mathcal{G}$, and then by algorithm from the point \textbf{[b]} of Theorem \ref{max_min_csp_det_theorem}  we obtain an assignment $\chi$ for $\mathcal{F}$.
  \par In case $\Opt(\mathcal{F})=0$ by claim \textbf{[b]} of Theorem \ref{max_min_csp_det_theorem} we have that $\Opt(\mathcal{G})=0$ as well. Therefore, since $\zeta$ gives us $\alpha$ approximation of $\Opt(\mathcal{G})$, we have that $\Val_{\zeta}(\mathcal{G})=0$. Finally,  by claim \textbf{[a]} of Theorem \ref{max_min_csp_det_theorem}  we have that $\Val_{\chi}(\mathcal{F}) \leq \delta/m$, which can be only possible if $\Val_{\chi}(\mathcal{F}) =0$.
  \par It remains to consider the case when $\Opt(\mathcal{F})\neq 0$, i.e.~$\Opt(\mathcal{F}) \geq 1/m$. In that case we have 
  \begin{equation}
    \begin{split}
      \frac{\Val_{\chi}(\mathcal{F})}{\Opt(\mathcal{F})} \leq     \frac{\Val_{\chi}(\mathcal{G})+\varepsilon}{\Opt(\mathcal{F})} \leq \frac{\Val_{\chi}(\mathcal{G})}{\Opt(\mathcal{F})} + \frac{\varepsilon}{\Opt(\mathcal{F})} \\
      \leq \frac{\Val_{\chi}(\mathcal{G})}{\Opt(\mathcal{G})} + \frac{\delta/m}{1/m}  \leq \alpha + \delta,
    \end{split}
  \end{equation}
  which finishes the proof of Theorem \ref{main_theorem_introduction_min}. 

\section{Conclusion and Some Open Questions} \label{conclusion_section}
	In this paper we introduced a reduction which shows how approximation algorithms working on regular unweighted instances of optimization CSPs can be converted (with an arbitrary small loss in approximation ratio) into approximation algorithms for weighted CSPs in which regularity is not imposed. One interesting question would be to see if we could use this result to obtain better approximation algorithms for different CSPs. Also, the aim of quantifying what makes the problems hard is interesting in its own right, and therefore it would be valuable to analyze whether some additional structure of CSP instances can always be assumed when studying their inapproximability.
 \par 
  It is not uncommon that reductions showing hardness of approximation output instances which satisfy some form of regularity. This work shows that we can not hope to obtain stronger inapproximability results by considering irregular instances of CSPs. However, for many other problems it is still not known whether regular instances might be easier to approximate; answering this question could facilitate search for optimal algorithms. One family of problems for which this is especially interesting topic due to their generality and applicability is defined as ``Max Ones'' in \cite{DBLP:journals/siamcomp/KhannaSTW00}. 
  \par 
  On the other side, let us remark that using irregular instances can also be instrumental for showing strong hardness results for certain problems, as recently shown in \cite{DBLP:conf/approx/AustrinS19} which treated some cardinality constrained CSPs, i.e. a variant of a CSP problem where we also prescribe the cardinality of zeros/ones in admissible assignments. Hence, it would be interesting to explore whether we can obtain better hardness results by considering more irregular/asymmetric instances for some problems for which satisfactory understanding of approximability is lacking. 
 \par 
\section*{Acknowledgments}
I am indebted to Per Austrin for pointing out the reduction in \cite{DBLP:conf/stoc/Trevisan01} to me. I also thank Johan H\r{a}stad for numerous useful comments which significantly improved the quality of presentation of this work.

  \bibliography{bibl}{}
  \bibliographystyle{siam}
    
  \appendix
  \setcounter{secnumdepth}{0}
  \section{Appendix}
  We state here concentration inequalities which give bounds on probability that certain random variable deviates from its mean. While these bounds are widely known, the form in which they appear can vary, and therefore we fix below the versions which are used in this paper.
  \par
  We use the following variant of Chernoff's inequality.
  \begin{lemma}\label{Chernoff2}
    Let $X=\sum_{i=1}^K X_i$, where $\{X_i\}_{i=1}^K$ are mutually independent random variables with range $\{0,1\}$. Then
    \begin{equation*}
      \Prob[ |X-\E[X] | \geq \delta \E[X]  ] \leq 2 e^{-\E[X] \delta^2 /4}, \quad \delta \in (0,1).
    \end{equation*}
  \end{lemma}
  Proof of this lemma can be found in \cite[Corollary~A.15]{DBLP:books/daglib/0023084}. Sometimes it will be more convenient to use the following corollary of the previous lemma.
  \begin{corollary}\label{Chernoff1}
    Let $X=\sum_{i=1}^K X_i$, where $\{X_i\}_{i=1}^K$ are mutually independent random variables with range $\{0,1\}$. Then
    \begin{equation*}
      \Pr\left[ \left|X - \E\left[X \right] \right| \geq \varepsilon K \right ] \leq 2 e^{-\frac{\varepsilon^2K}{4}}.
    \end{equation*}
    \begin{proof}
      To proof follows by using inequality from Lemma $\ref{Chernoff2}$ with $\delta = \varepsilon K / \E[X]$, and noting that $K/\E[X] \geq 1$.
    \end{proof}
  \end{corollary}
  We also need a concentration bound for sum of random variables with range $[0,b], b \in \mathbb{R}$. For that, we use the following variant of Hoeffding's inequality \cite{doi:10.1080/01621459.1963.10500830}.
  \begin{lemma}\label{hoeff_bound}
    Let $X_1,\hdots,X_K$ be independent variables such that range of each $X_i$ is $[0,b]$, where $b \in \mathbb{R}$. Then for $X=\sum_{i=1}^k X_i$ we have 
    \begin{equation*}
      \Prob[ |X-\E[X] | \geq t ] \leq 2 e^{-\frac{t^2}{K b^2}}.
    \end{equation*}
  \end{lemma}
\end{document}

%% file: structure.tex
\usepackage[dvipsnames]{xcolor}
\usepackage[T1]{fontenc} 

\usepackage[utf8]{inputenc} 

\usepackage{graphicx} 
\graphicspath{{Figures/}} 

\usepackage{enumitem} 

\usepackage{lipsum} 

\usepackage{subfig} 
\usepackage{amsmath,amssymb,amsthm} 

\usepackage{varioref} 

\usepackage{dsfont}
\usepackage{bm}
\usepackage[titletoc]{appendix}

\usepackage[margin=32mm]{geometry}

\usepackage[hidelinks]{hyperref} 

\usepackage{multirow,bigdelim}

\newtheorem{same-number}{Theorem}
\theoremstyle{definition} 
\newtheorem{definition}[same-number]{Definition}

\theoremstyle{lemma} 
\newtheorem{lemma}[same-number]{Lemma}

\theoremstyle{corollary} 
\newtheorem{corollary}[same-number]{Corollary}

\theoremstyle{plain} 
\newtheorem{theorem}[same-number]{Theorem}

\theoremstyle{remark} 

